\documentclass[conference]{IEEEtran}

\usepackage{tikz}
\usetikzlibrary{arrows.meta,positioning,fit,calc,backgrounds}

\IEEEoverridecommandlockouts

\usepackage{cite}
\usepackage{multirow}
\usepackage{amsmath,amssymb,amsfonts}
\usepackage{algorithmic}
\usepackage{graphicx}
\usepackage{textcomp}
\usepackage{xcolor}
\def\BibTeX{{\rm B\kern-.05em{\sc i\kern-.025em b}\kern-.08em
    T\kern-.1667em\lower.7ex\hbox{E}\kern-.125emX}}

\graphicspath{{image/}}

\begin{document}

\title{Soft Graph Diffusion Transformer for \\ MIMO Detection
}

\author{
\IEEEauthorblockN{Nan~Jiang$^*$,~Jiadong~Hong$^*$,~Lei~Liu$^*$,~Xinyu~Bian$^\dagger$,~Wenjie~Wang$^\dagger$,~Zhaoyang~Zhang$^*$}
\IEEEauthorblockA{$^*$College of Information Science and Electronic Engineering, Zhejiang University, Hangzhou, China\\
$\dagger$Theory Lab, Central Research Institute, 2012 Labs, Huawei Technologies Co., Ltd., Hong Kong, China\\
Email: \{nan\_jiang, jiadong5, lei\_liu, zhzy\}@zju.edu.cn, \{bian.xinyu, wang.wenjie\}@huawei.com}
}

\maketitle
\bstctlcite{IEEEexample:BSTcontrol}

% \begin{abstract}
% Learning-based MIMO detection has recently attracted significant attention due to its strong empirical performance. In this paper, we propose the \emph{Soft Graph Diffusion Transformer (SGDiT)}, a conditional flow matching framework that reformulates MIMO detection as a progressive denoising process. Specifically, symbol recovery is modeled as a progressive denoising trajectory that starts from Gaussian initial states and is progressively denoised toward the posterior estimate conditioned on channel observations. An adaptive layer normalization (AdaLN)-conditioned soft graph transformer is employed to parameterize the denoising dynamics, enabling stage-aware information integration between observation and symbol domains. To better align with the discrete nature of symbol detection, we further adopt a cross-entropy-based training objective that directly models bit-wise posterior probabilities, providing a more suitable inductive bias than conventional regression-based formulations. Experimental results across various MIMO system configurations demonstrate that SGDiT achieves competitive bit error rate (BER) performance compared with representative baselines. Furthermore, the proposed model exhibits good generalization capability across different channel conditions. Overall, the SGDiT framework provides an effective and practical approach for neural MIMO detection.
% \end{abstract}

\begin{abstract}
Learning-based MIMO detection has shown strong empirical performance, yet existing methods typically rely on fixed-depth architectures without explicitly modeling the progressive refinement of symbol estimates. In this paper, we revisit MIMO detection from a flow matching perspective and propose the Soft Graph Diffusion Transformer (SGDiT), which reformulates detection as a noise-level-conditioned denoising process that progressively transforms a Gaussian initialization toward the posterior conditioned on channel observations.  An adaptive layer normalization (AdaLN)-conditioned soft graph transformer is employed to parameterize the denoising dynamics, enabling stage-aware information integration between observation and symbol domains. To better align with the discrete nature of symbol detection, we further adopt a cross-entropy-based training objective that directly models bit-wise posterior probabilities, providing a more suitable inductive bias than conventional regression-based formulations. Experimental results across various MIMO system configurations demonstrate that SGDiT achieves competitive bit error rate (BER) performance compared with representative baselines. Furthermore, the proposed model exhibits good generalization capability across different channel conditions. Overall, the SGDiT framework provides an effective and practical approach for neural MIMO detection.
\end{abstract}

% \begin{IEEEkeywords}
% MIMO detection, conditional flow matching, diffusion transformer, channel generalization.
% \end{IEEEkeywords}


\section{Introduction}

Multiple-input multiple-output (MIMO) detection is a central problem in modern wireless communication systems. While maximum-likelihood (ML) detection \cite{verdu1998multiuser} achieves optimal performance, its computational complexity grows rapidly with system dimension, making it impractical for large-scale systems. To address this issue, iterative Bayesian inference algorithms, including approximate message passing (AMP) \cite{amp}, orthogonal approximate message passing (OAMP) \cite{oamp}, vector approximate message passing (VAMP) \cite{vamp}, and memory approximate message passing (MAMP) \cite{memoryamp}, have been widely adopted. More recently, model-driven deep learning approaches incorporate learnable parameters into these algorithms, such as DetNet \cite{detnet}, OAMP-Net \cite{oampnet}, and its improved variant OAMPNet2 \cite{oampnet2}. 

In parallel, learning-based MIMO detectors explore flexible neural architectures for iterative denoising. Graph-based and message-passing inspired models, such as AMP-based graph neural networks (AMP-GNN) \cite{ampgnn}, enhance information exchange across symbols. Recurrent and attention-based architectures have also been investigated, including RE-MIMO \cite{remimo} and Transformer-based MIMO detectors \cite{transformer_mimo, transformer_comm} and related models \cite{ecc_transformer}. Among these, the Soft Graph Transformer (SGT) \cite{sgt} provides a structured neural detection framework incorporating the MIMO factor graph. However, its standard training paradigm lacks an explicit mechanism for progressive denoising, which may limit performance under severe noise.

% To address this limitation, we revisit MIMO detection from a flow matching perspective. Flow matching \cite{flowmatching} provides a principled framework for modeling progressive denoising processes, as demonstrated by diffusion models and diffusion transformers \cite{dit}. This motivates us to integrate flow matching with the SGT architecture to model denoising trajectories guided by channel observations. When adapting flow matching models to MIMO detection, the training objective becomes critical. For Quadrature Phase Shift Keying (QPSK), real-valued MIMO detection can be formulated as a binary signal recovery problem, where the choice of prediction parameterization and loss function plays a key role. Recent studies on binary flow matching \cite{binaryfm} show that signal-space prediction provides more stable optimization, while binary cross-entropy naturally matches the bit-wise posterior, whereas mean squared error may introduce unnecessary geometric constraints \cite{binaryfm}. These observations guide the design of flow matching detectors for MIMO systems. Recently, large-scale models and embodied intelligence have emerged as key enablers for 6G integrated systems \cite{li2025large}, motivating more flexible and adaptive signal processing frameworks.

To address this limitation, we revisit MIMO detection from a flow matching perspective. Flow matching \cite{flowmatching} provides a principled framework for modeling progressive denoising processes, as demonstrated by diffusion models and diffusion transformers \cite{dit}. This motivates us to integrate flow matching with the SGT architecture to model denoising trajectories guided by channel observations. When applied to MIMO detection, the training objective becomes critical. For QPSK, real-valued MIMO detection can be viewed as a binary signal recovery problem, where the choice of prediction parameterization and loss function is crucial. Recent studies on binary flow matching \cite{binaryfm} show that signal-space prediction leads to more stable optimization, while binary cross-entropy better matches the bit-wise posterior compared with mean squared error \cite{binaryfm}. Recently, large-scale models and embodied intelligence have emerged as key enablers for 6G integrated systems \cite{li2025large}, motivating the need for more flexible and adaptive signal processing frameworks.

Motivated by the above, we propose the \emph{Soft Graph Diffusion Transformer (SGDiT)} for MIMO detection. The proposed method builds upon the SGT backbone and introduces adaptive layer normalization (AdaLN) to condition the denoising process across different stages. Under this formulation, the detector performs a sequence of conditional denoising steps from a Gaussian initial state toward the transmitted symbol estimate. During training, we adopt a cross-entropy objective aligned with the binary structure of the detection problem, while inference requires only a small number of denoising steps.

The main contributions of this paper are summarized as follows:
\begin{itemize}
\item We propose a novel flow matching perspective for MIMO detection by reformulating it as a continuous conditional denoising process. To realize this, we develop the Soft Graph Diffusion Transformer (SGDiT), which integrates the SGT architecture with AdaLN-based stage conditioning to explicitly learn progressive denoising trajectories.

\item We introduce a prediction-loss alignment strategy tailored for the discrete nature of MIMO systems. By adopting a signal-space training objective based on binary cross-entropy, our framework avoids the geometric constraints of conventional regression losses and provides a superior inductive bias for bit-wise posterior estimation.
\end{itemize}

Through extensive evaluations on various MIMO systems under different channel settings, we validate the effectiveness of the proposed framework. Experimental results demonstrate that SGDiT achieves competitive bit error rate (BER) performance compared with representative baselines.

\section{Preliminaries}

\subsection{MIMO System Model}

We focus on a standard MIMO communication architecture, operating under the assumption that perfect channel state information (CSI) is available at the receiver. At the transmitter side, the raw information bits are processed through an error correction code and subsequently mapped into complex-valued constellation symbols. These symbols are then allocated across a time-frequency resource grid and emitted via multiple transmit antennas. Upon reception, the system can be effectively modeled as a stacked MIMO channel. In the complex domain, the observation vector is formulated as
\begin{equation}
\mathbf{y}_c = \mathbf{H}_c\mathbf{x}_c + \mathbf{n}_c,
\end{equation}
where $\mathbf{y}_c \in \mathbb{C}^{N_r}$ represents the received signal, $\mathbf{x}_c \in \mathbb{C}^{N_t}$ is the transmitted symbol vector, $\mathbf{H}_c \in \mathbb{C}^{N_r \times N_t}$ denotes the complex channel matrix, and $\mathbf{n}_c \sim \mathcal{CN}(\mathbf{0}, \sigma_c^2 \mathbf{I})$ accounts for the additive white Gaussian noise (AWGN).

To facilitate algorithm design and subsequent theoretical analysis, we transform this complex model into an equivalent \emph{real-valued representation}. By systematically decoupling the real and imaginary components, the linear system is reformulated as
\begin{equation}
\underbrace{
\begin{bmatrix}
\Re(\mathbf{y}_c) \\
\Im(\mathbf{y}_c)
\end{bmatrix}
}_{\mathbf{y} \in \mathbb{R}^{2N_r}}
=
\underbrace{
\begin{bmatrix}
\Re(\mathbf{H}_c) & -\Im(\mathbf{H}_c) \\
\Im(\mathbf{H}_c) & \Re(\mathbf{H}_c)
\end{bmatrix}
}_{\mathbf{H} \in \mathbb{R}^{2N_r \times 2N_t}}
\underbrace{
\begin{bmatrix}
\Re(\mathbf{x}_c) \\
\Im(\mathbf{x}_c)
\end{bmatrix}
}_{\mathbf{x} \in \mathbb{R}^{2N_t}}
+
\underbrace{
\begin{bmatrix}
\Re(\mathbf{n}_c) \\
\Im(\mathbf{n}_c)
\end{bmatrix}
}_{\mathbf{n} \in \mathbb{R}^{2N_r}},
\end{equation}

where $\mathbf{n} \sim \mathcal{N}(\mathbf{0}, \frac{\sigma_c^2}{2} \mathbf{I}) = \mathcal{N}(\mathbf{0}, \mathbf{\Sigma})$ is real-valued Gaussian noise, $\mathbf{\Sigma} = \mathrm{diag}(\sigma_1^2, \ldots, \sigma_{2N_r}^2)$. The remainder of our algorithm design relies heavily on this real-valued formulation.

In typical receiver architectures, a linear equalizer is initially employed to compute coarse estimates of the transmitted symbols $\mathbf{x}$ subject to the observation constraint $\mathbf{y} = \mathbf{H}\mathbf{x}$. Following this, a symbol demapper performs Bayesian denoising, utilizing the discrete prior distribution of $\mathbf{x}$ to accomplish the symbol-to-bit conversion. Within the approximate message passing (AMP) framework \cite{amp}, detection can be interpreted as iterative message passing on a bipartite factor graph, where observation nodes enforce the linear constraint $\mathbf{y}=\mathbf{H}\mathbf{x}$ and variable nodes encode symbol priors. The receiver progressively refines symbol estimates through iterative information exchange.

\subsection{Soft Graph Transformer (SGT)}

The Soft Graph Transformer (SGT) \cite{sgt} is a neural detection architecture designed for soft-input soft-output (SISO) inference.  To align the MIMO detection problem with attention-based models, SGT employs a graph-aware tokenization scheme that maps the linear system onto a bipartite representation. Specifically, the factor graph is partitioned into two subgraphs. The observation nodes are tokenized into linear-constraint tokens, denoted as $\mathcal{T}_{\mathrm{lin}} = \{ \tau_j = (y_j, \mathbf{h}_j, \sigma_j^2) \}$, where the channel information is incorporated into the token features to explicitly encode local likelihood constraints. The variable nodes are represented as symbolic tokens $\mathcal{T}_{\mathrm{sym}} = \{ x_i \}$, corresponding to the transmitted symbols and their associated soft priors.

SGT utilizes both self-attention and cross-attention to process these tokens. Self-attention captures contextual dependencies within each homogeneous token set, while cross-attention \cite{crossmpt} enables information exchange between observation and symbol domains, similar to recent transformer-based message passing architectures. By combining these mechanisms, SGT integrates contextual encoding with structured message passing in a unified architecture. Furthermore, SGT provides a well-defined SISO interface that accepts prior symbol beliefs and channel observations, maps them into a shared latent space, and processes them through stacked Transformer layers. The refined symbolic representations are then projected to bit-wise posterior log-likelihood ratios (LLRs). 

While SGT provides an effective mechanism to incorporate prior information through structured token interactions, the progressive denoising process of noisy intermediate estimates is not explicitly modeled; rather, it is merely implicitly learned through the fixed stacked layers. This structural limitation directly motivates the denoising formulation adopted in our SGDiT framework, where the continuous progression of intermediate signal estimates is explicitly parameterized and learned across different continuous stages.

\subsection{Flow Matching}

Flow matching provides a general framework for constructing denoising processes that gradually transform a simple prior distribution into a target data distribution. Instead of directly mapping from input to output in a single step, this framework models the estimation process as a sequence of intermediate updates, starting from a noisy initial state and progressively denoising it toward the desired signal. From this perspective, a neural network is trained to capture how the current estimate should be updated at different stages of the denoising process. This view aligns well with the iterative structure of classical MIMO detection algorithms. While originally developed for continuous-valued data, recent work has explored extending this framework to discrete or binary settings. In such cases, the choice of prediction parameterization and training objective becomes particularly important.

In particular, recent studies on binary flow matching \cite{binaryfm} have shown that directly supervising the induced velocity using regression objectives, such as mean squared error (MSE), may lead to a mismatch between the prediction space and the loss space when the target variables are discrete. Instead, defining the objective directly in the signal prediction space provides a formulation that is better aligned with the underlying discrete structure.

These observations are directly relevant to MIMO detection. For commonly used modulation schemes, the real-valued decomposition leads to a structured discrete signal recovery problem, where symbols are drawn from a finite set. Motivated by this connection, we formulate MIMO detection as a conditional denoising process driven by the channel observations $(\mathbf{y}, \mathbf{H})$, and adopt a training objective that is consistent with the discrete structure of the signal. The corresponding model design and training formulation are described in the next section.

\section{Methodology}

In this section, we present the proposed Soft Graph Diffusion Transformer (SGDiT) framework for MIMO detection. Standard iterative detectors typically operate through a fixed number of discrete updates, which may not fully capture the progressive denoising trajectory, especially under severe channel noise. To address this, our methodology formulates the detection process within a flow matching framework. Specifically, we first introduce a noise-level-conditioned architecture to parameterize the denoising trajectory, followed by a prediction-loss alignment strategy designed to handle the discrete nature of symbol recovery.

\begin{figure*}[t]
\centerline{\includegraphics[width=\textwidth]{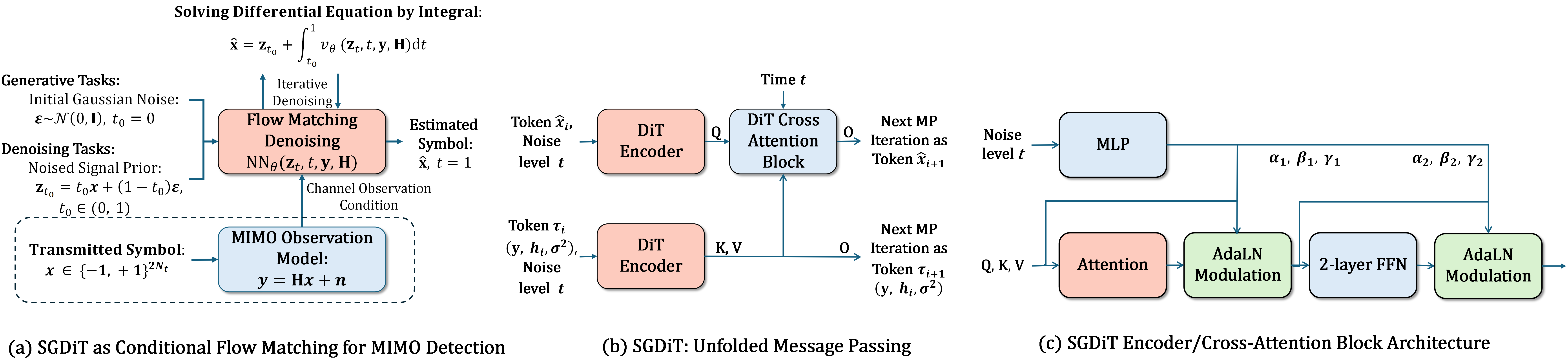}}
\caption{Detailed architecture of the proposed Soft Graph Diffusion Transformer (SGDiT). 
\textbf{(a) SGDiT as Conditional Flow Matching for MIMO Detection:} The detection process is formulated as a progressive denoising process. 
Starting from an initial noisy state $t_0$, the model leverages channel observations $(\mathbf{y}, \mathbf{H})$ to iteratively denoise the latent variable via integration. \textbf{(b) Unfolded Message Passing:} At a specific noise level $t$, the symbol tokens $\hat{\mathbf{x}}_i$ and linear-constraint tokens $\tau_i$ are processed by diffusion transformer (DiT) encoders. Cross-attention is applied to aggregate observation-guided updates.  \textbf{(c) SGDiT Encoder/Cross-Attention Block Architecture:} The noise level $t$ is embedded via a multi-layer perceptron (MLP) to generate modulation parameters $(\alpha, \beta, \gamma)$, enabling adaptive layer normalization (AdaLN) for noise-level-aware feature extraction.}
\label{fig:SGDiT_framework}

\end{figure*}

\subsection{Noise-Level-Conditioned Architecture}

While standard iterative detectors rely on layer-wise discrete mappings, flow matching essentially formulates signal estimation as a progressive denoising process. To properly parameterize this process within the SGT backbone, the network must be explicitly conditioned on the noise level variable $t$.

\textbf{Conditional Flow Matching:}
Instead of mapping the initial estimate to the final output through discrete steps, we formulate a progressive denoising trajectory. By introducing an intermediate latent variable $\mathbf{z}_t$, we construct a probability path that continuously interpolates between a Gaussian prior and the target clean signal $\mathbf{x}$. The forward trajectory at noise level $t \in [0,1]$ is defined as:
\begin{equation}
\mathbf{z}_t = t\mathbf{x} + (1-t)\boldsymbol{\epsilon}, \quad \boldsymbol{\epsilon}\sim\mathcal{N}(\mathbf{0},\mathbf{I}),
\end{equation}
where $\boldsymbol{\epsilon}$ represents standard Gaussian noise. For $t=0$, $\mathbf{z}_0$ is purely noise, whereas for $t=1$, $\mathbf{z}_1$ recovers the transmitted signal $\mathbf{x}$. Under this formulation, MIMO detection is cast as a conditional denoising task, where the network progressively denoises $\mathbf{z}_t$ toward the clean signal $\mathbf{x}$, guided by the channel observations $(\mathbf{y}, \mathbf{H})$.

\textbf{Noise-Level-Conditioned SGDiT Backbone:} 
To seamlessly embed this denoising dynamics into the network, we extend the standard SGT architecture into SGDiT by incorporating a dynamic noise-level conditioning mechanism. As illustrated in Fig.~\ref{fig:SGDiT_framework}(b), the system state at noise level $t$ is explicitly tokenized into symbol tokens $\hat{\mathbf{x}}_i$ and linear-constraint tokens $\tau_i = (y, h_i, \sigma^2)$.

Crucially, rather than merely appending $t$ as a static input feature, we modulate the transformer backbone using an adaptive layer normalization (AdaLN) mechanism following the design of DiT~\cite{dit}. As depicted in Fig.~\ref{fig:SGDiT_framework}(c), the scalar noise level $t$ is first embedded and passed through a multi-layer perceptron (MLP) to generate dimension-wise modulation parameters, including residual scaling $\boldsymbol{\alpha}(t)$, normalization scale $\boldsymbol{\gamma}(t)$, and shift $\boldsymbol{\beta}(t)$.

Let $\mathbf{u}$ denote the input token representations. Following the DiT design~\cite{dit}, the transformer block is modulated by noise-level-dependent parameters generated from an MLP. Specifically, the noise-level embedding produces a set of modulation parameters $\{\boldsymbol{\alpha}_1(t), \boldsymbol{\alpha}_2(t), \boldsymbol{\gamma}_1(t), \boldsymbol{\beta}_1(t), \boldsymbol{\gamma}_2(t), \boldsymbol{\beta}_2(t)\}$, which are used to condition both the attention and feed-forward branches.

The attention branch is formulated as
\begin{equation}
\mathbf{h} = \mathbf{u} + \boldsymbol{\alpha}_1(t) \odot \mathrm{Attn}\big(\boldsymbol{\gamma}_1(t) \odot \mathrm{LN}(\mathbf{u}) + \boldsymbol{\beta}_1(t)\big),
\end{equation}

and the feed-forward branch is given by
\begin{equation}
\mathbf{y} = \mathbf{h} + \boldsymbol{\alpha}_2(t) \odot \mathrm{FFN}\big(\boldsymbol{\gamma}_2(t) \odot \mathrm{LN}(\mathbf{h}) + \boldsymbol{\beta}_2(t)\big).
\end{equation}

where $\boldsymbol{\alpha}(t)$, $\boldsymbol{\gamma}(t)$, and $\boldsymbol{\beta}(t)$ are dynamically generated from the noise-level embedding. This mechanism enables noise-level-dependent feature modulation, allowing the network to adaptively adjust its contextual encoding and message-passing behavior across different denoising stages.

\subsection{Prediction-Loss Space Alignment for Discrete MIMO}

While the SGDiT architecture effectively parameterizes the progressive denoising trajectory, the training objective must be carefully designed to accommodate the discrete nature of MIMO signals. In modern MIMO systems, the target symbols are inherently drawn from a discrete finite alphabet. Through standard bit-mapping, the real-valued target vector $\mathbf{x}$ can be equivalently formulated in a strictly binary space, i.e., $\mathbf{x}\in\{-1,+1\}^{2N_t}$. 

\textbf{Signal-Space Prediction:} 
Standard flow matching models typically predict the induced velocity field $\mathbf{v}_t = \mathbf{x} - \boldsymbol{\epsilon}$. However, directly predicting this continuous velocity requires the network to learn unbounded regression targets, which creates an optimization mismatch with the bounded, discrete structure of the symbols. To resolve this mismatch, we configure the neural network $\mathrm{NN}_\theta(\mathbf{z}_t, t, \mathbf{y}, \mathbf{H})$ to operate exclusively under signal-space prediction. Instead of predicting the velocity, the model directly outputs the estimated clean signal $\hat{\mathbf{x}}_{\theta}$. 
The corresponding denoising direction (velocity field) is then implicitly derived as
\begin{equation}
\mathbf{v}_{\theta} = \frac{\hat{\mathbf{x}}_{\theta} - \mathbf{z}_t}{1 - t}.
\end{equation}

\textbf{Cross-Entropy Loss:} 
Under signal-space prediction, the training objective should be defined in the same space as the prediction target to avoid prediction--loss mismatch. 
For discrete MIMO detection, the transmitted symbols after real-valued decomposition follow an i.i.d. binary prior, i.e., $\mathbf{x} \in \{-1,+1\}^{2N_t}$. 
This structure naturally induces a factorized probabilistic model over individual bits.

Specifically, we map the binary signal into label space as
\begin{equation}
\mathbf{b} = (\mathbf{x}+1)/2 \in \{0,1\}^{2N_t},
\end{equation}
and interpret the network output $\hat{\mathbf{x}}_{\theta}$ as logits of independent Bernoulli variables. 
The corresponding bit-wise probabilities are given by
\begin{equation}
p_{\theta,i} = \sigma(\hat{x}_{\theta,i}), \quad i=1,\ldots,2N_t,
\end{equation}
where $\sigma(\cdot)$ denotes the sigmoid function.

Under the conditional independence assumption, the likelihood factorizes as
\begin{equation}
p(\mathbf{b} \mid \hat{\mathbf{x}}_{\theta}) = \prod_{i=1}^{2N_t} p_{\theta,i}^{b_i} (1 - p_{\theta,i})^{1-b_i}.
\end{equation}

Maximizing this likelihood is equivalent to minimizing the cross-entropy loss
\begin{equation}
\mathcal{L}_{\mathrm{CE}} 
= 
-\mathbb{E}_{t, \mathbf{x}, \boldsymbol{\epsilon}}
\left[
\sum_{i=1}^{2N_t} 
b_i \log p_{\theta,i} 
+ 
(1-b_i)\log(1-p_{\theta,i})
\right].
\end{equation}

This formulation directly follows from the i.i.d. binary structure of the MIMO detection problem and provides a probabilistically consistent objective for symbol-wise inference. 
Compared with regression-based objectives such as MSE, which implicitly impose a continuous Euclidean geometry, the cross-entropy loss aligns with the discrete signal topology and yields a more appropriate inductive bias for bit-wise posterior estimation. 
This observation is consistent with recent findings on prediction--loss space alignment for binary flow matching, where probabilistic objectives are shown to be more suitable for independent symbolic recovery tasks.

At inference time, the detector starts from a Gaussian initialization and iteratively updates the estimate using a small number of discretized denoising steps. 
After the final step, a hard decision is applied to obtain the transmitted symbols.

\section{Experiment}

In this section, we evaluate the proposed SGDiT detector from three aspects: 
1) an ablation study to validate the proposed prediction-loss alignment strategy; 
and 
2) comparison with representative baseline methods under matched Rayleigh fading channels; 
3) evaluation of generalization performance under channel distribution shift using the 3GPP UMa model.

\subsection{Simulation Setup}

We consider MIMO systems with $N_t = N_r \in \{8,16\}$ under QPSK modulation. For the distribution shift setting, we adopt the 3GPP UMa channel model, which introduces spatial correlation and large-scale fading effects. The signal-to-noise ratio (SNR) is defined as $E_b/N_0$. Performance is evaluated using bit error rate (BER), averaged over multiple independent channel realizations. In training, the noise level $t$ is sampled from a uniform distribution over $[0,1]$, i.e., $t \sim \mathcal{U}(0,1)$. To avoid numerical instability near the terminal point, we clip the maximum value of $t$ to $0.99$ during training. This strategy ensures stable optimization while preserving the effectiveness of the continuous denoising formulation. At inference, SGDiT performs $K$ discretized denoising steps to approximate the continuous flow. Unless otherwise specified, we set $K=2$ in the BER experiments. We compare SGDiT with both model-based and learning-based baselines, including LMMSE, OAMP, OAMPNet2, and SGT. 
For the $8\times8$ system, the maximum likelihood (ML) detector is included as a reference upper bound.

\subsection{Ablation Study}

We conduct an ablation study to evaluate the proposed prediction-loss alignment strategy. All variants share the same SGDiT backbone, while the prediction parameterization and loss function are modified. Specifically, we compare: 
1) signal-space prediction with BCE loss (proposed), 
2) signal-space prediction with MSE loss, 
3) signal-space prediction with velocity MSE supervision, 
and 4) velocity prediction with MSE loss.

\begin{figure}[t]
\centering
\includegraphics[width=\columnwidth]{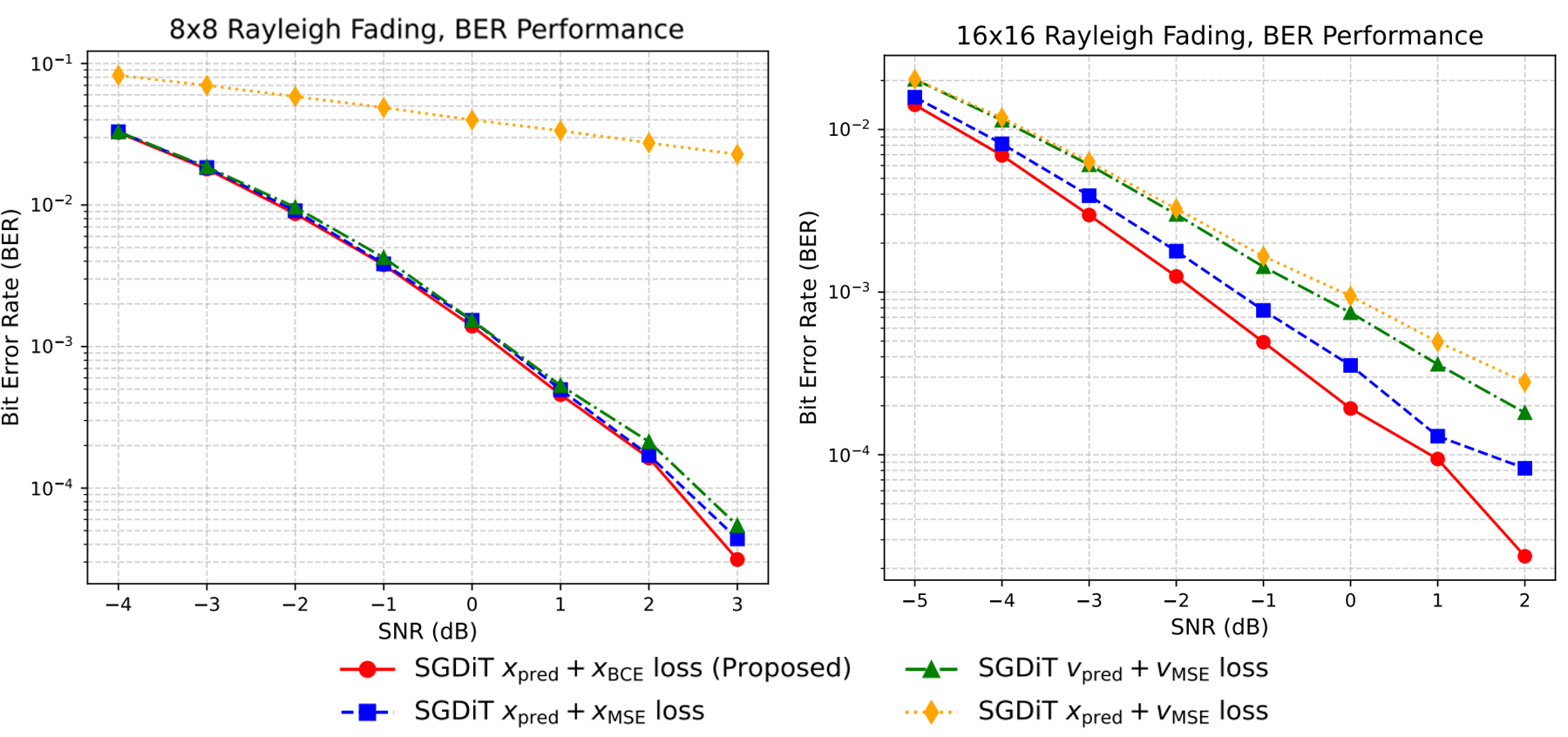}
\caption{Ablation study on prediction-loss alignment under Rayleigh fading channels.}
\label{fig:exp2_ablation}
\end{figure}

As shown in Fig.~\ref{fig:exp2_ablation}, the proposed BCE-based signal-space formulation achieves the lowest BER across both system sizes. In contrast, MSE-based and velocity-based variants exhibit noticeable degradation. While signal-space prediction with MSE still benefits from directly modeling the target signal, the regression objective does not fully capture the discrete nature of the symbols. Velocity-based supervision further degrades performance, suggesting that misalignment between the prediction target and the detection objective increases optimization difficulty. These results highlight the importance of aligning the prediction parameterization with the training objective. For discrete signal detection, combining signal-space prediction with a classification-based BCE loss provides a better inductive bias, leading to improved estimation accuracy.

Overall, the ablation results confirm that prediction-loss alignment is critical for effective flow-based MIMO detection.

\subsection{Baseline Comparisons under Rayleigh Channels}

We first evaluate all detectors under matched Rayleigh fading channels. The goal is to examine whether the proposed denoising formulation can achieve competitive performance when training and testing channel distributions are consistent. We consider both $8\times8$ and $16\times16$ MIMO systems and compare SGDiT with classical model-based and learning-based baselines.

\begin{figure}[t]
\centering
\includegraphics[width=\columnwidth]{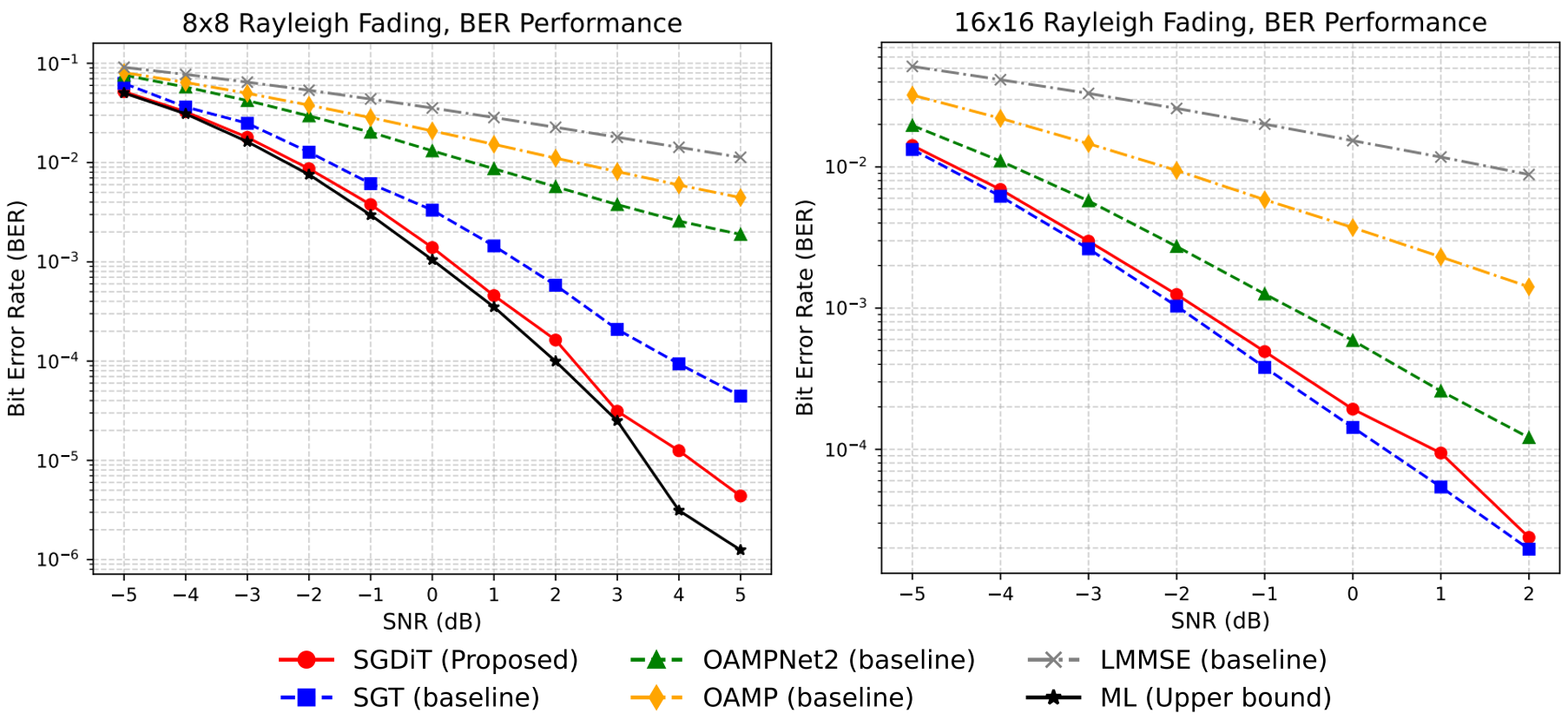}
\caption{BER performance comparison between SGDiT and baseline detectors under Rayleigh fading channels for $8\times8$ and $16\times16$ MIMO systems.}
\label{fig:exp1_baselines}
\end{figure}

The results are shown in Fig.~\ref{fig:exp1_baselines}. For the $8\times8$ system, SGDiT achieves the best performance across the entire SNR range and closely approaches the ML reference. Compared with model-based algorithms such as OAMP and LMMSE, SGDiT provides a clear performance gain, indicating the effectiveness of the progressive denoising formulation. Compared with learning-based baselines, SGDiT also shows consistent improvements over OAMPNet2 and SGT. While these methods rely on fixed-depth architectures, SGDiT explicitly models the denoising trajectory, enabling stage-adaptive updates and more accurate symbol estimation. For the $16\times16$ system, SGDiT continues to outperform OAMP, OAMPNet2, and LMMSE across all SNR values. Compared with SGT, the performance becomes comparable, with SGT achieving slightly lower error rates in certain SNR regions. This suggests that, as system dimension increases, optimizing the denoising trajectory becomes more challenging, while fixed-depth mappings may provide more stable approximations under limited model capacity. Nevertheless, the gap between SGDiT and classical baselines remains consistent, demonstrating scalability with system dimension.

Overall, these results show that SGDiT achieves competitive performance across different system sizes and that the proposed denoising formulation provides a clear improvement over both model-based and existing learning-based detectors.

\subsection{Generalization Capability Evaluation}

Finally, we evaluate all detectors under channel distribution shift. The models are trained on Rayleigh fading channels and directly tested on 3GPP UMa channels without retraining. This experiment examines whether the learned detector can maintain competitive performance when the testing channel differs from the training distribution.

\begin{figure}[t]
\centering
\includegraphics[width=\columnwidth]{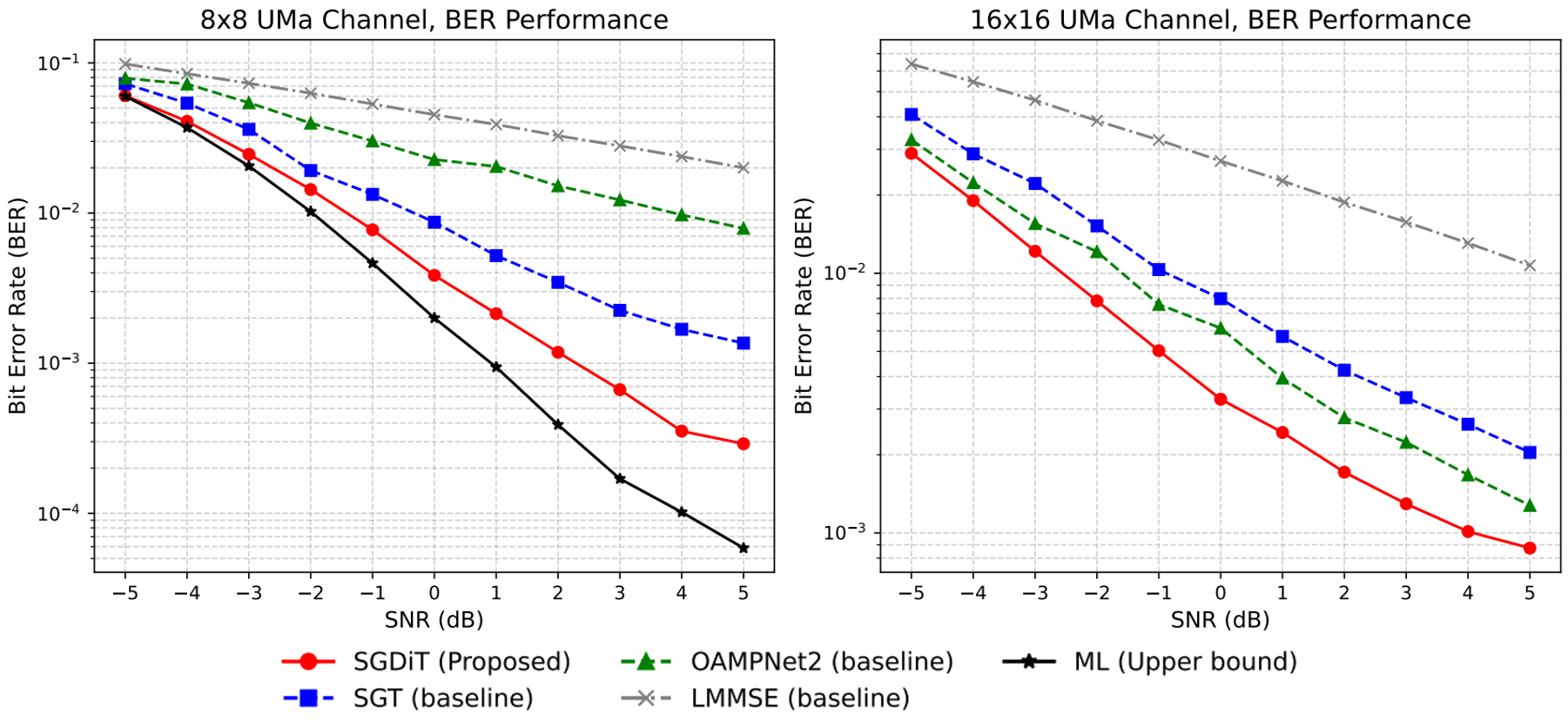}
\caption{Performance under channel distribution shift. Models are trained on Rayleigh fading channels and directly evaluated on 3GPP UMa channels without retraining.}
\label{fig:exp3_uma}
\end{figure}

The results are shown in Fig.~\ref{fig:exp3_uma}. All methods experience performance degradation due to the change in channel statistics. Model-based algorithms such as OAMP are affected because their assumptions on channel distribution are no longer fully satisfied, and learning-based detectors also show varying degrees of performance loss. Compared with the baselines, SGDiT maintains a consistent performance level and preserves a relative advantage across most SNR regions. This suggests that the continuous denoising formulation, together with the structured SGT backbone, helps the model learn a transferable mapping from noisy observations to symbol estimates.

\subsection{Computational Complexity Analysis}

\textbf{A. Theoretical Complexity Discussion}

We provide a qualitative complexity comparison of SGDiT with representative baselines. SGDiT performs detection via $K$ denoising steps, each corresponding to one forward pass of the SGT backbone. As a result, the overall complexity scales linearly with $K$ and can be viewed as $K$ times that of a standard SGT inference. A detailed comparison is summarized in Table~\ref{tab:complexity}.

\begin{table}[t]
\centering
\caption{Computational Complexity Comparison}
\begin{tabular}{l c}
\hline
Method & Complexity \\
\hline
ML Detection & $O(M^{N_t})$ \\
OAMP / OAMPNet & $O(K N_r N_t^2)$ \\
Transformer-based MIMO & $O(N_r N_t^2 + L N_t^2 d_{\text{model}})$ \\
SGT & $O(L (N_r^2 + N_t^2 + N_r N_t)\, d_{\text{model}})$ \\
SGDiT (Ours) & $O(KL (N_r^2 + N_t^2 + N_r N_t)\, d_{\text{model}})$ \\
\hline
\end{tabular}
\label{tab:complexity}
\end{table}

Compared with classical methods, ML detection has exponential complexity in the number of transmit antennas, making it impractical for moderate and large-scale systems. Iterative algorithms such as OAMP and OAMPNet exhibit polynomial complexity, with cost increasing with both system dimension and iteration number. Transformer-based detectors, including SGT, introduce higher computational overhead due to attention operations, but benefit from parallelizable implementations. In this context, SGDiT introduces a controllable overhead through the number of denoising steps $K$. When $K$ is small, its complexity is comparable to a single-pass Transformer-based detector, while larger $K$ improves performance at the cost of increased computation, enabling a flexible trade-off between accuracy and efficiency.

\textbf{B. Empirical Runtime Evaluation}

We further report the measured inference runtime on a top-tier 24GB gaming GPU with a batch size of 1000.

\begin{table}[t]
\centering
\caption{Measured runtime per batch (seconds)}
\begin{tabular}{l cc}
\hline
Method & $8\times8$ & $16\times16$ \\
\hline
LMMSE & 0.0010 & 0.0013 \\
OAMP & 0.0026 & 0.0040 \\
OAMPNet2 & 0.0016 & 0.0016 \\
SGT & 0.0133 & 0.0273 \\
ML Detection & 2.1388 & / \\
\hline
SGDiT ($K=1$) & 0.0155 & 0.0289 \\
SGDiT ($K=3$) & 0.0488 & 0.0975 \\
SGDiT ($K=5$) & 0.0854 & 0.1689 \\
\hline
\end{tabular}
\label{tab:runtime}
\end{table}

As shown in Table~\ref{tab:runtime}, SGDiT with $K=1$ achieves runtime comparable to SGT, indicating that a single denoising step has a similar cost as one backbone forward pass. As $K$ increases, the runtime grows approximately linearly, which is consistent with the above analysis. Overall, the results demonstrate that SGDiT provides a controllable trade-off between computational complexity and detection performance, where a small number of denoising steps is sufficient to achieve strong performance.

\section{Conclusion}

In this paper, we proposed the Soft Graph Diffusion Transformer (SGDiT), a flow-matching-based framework that reformulates MIMO detection as a noise-level-conditioned denoising process. Built upon a graph-based SGT backbone, SGDiT incorporates adaptive layer normalization (AdaLN) to enable stage-aware modulation across denoising steps, and adopts a prediction–loss alignment strategy tailored to discrete signals via signal-space prediction with a binary cross-entropy objective. Experimental results show that SGDiT achieves competitive performance under matched Rayleigh fading and maintains strong generalization under channel distribution shifts. Overall, the proposed framework demonstrates that integrating noise-level-conditioned denoising with prediction–loss alignment provides an effective and practical design paradigm for neural MIMO detection. 

% The proposed framework provides an effective design paradigm for neural MIMO detection, and offers a promising direction for integrating flow-based modeling with structured architectures in future communication systems.

% \section{Conclusion}

% In this paper, we proposed the Soft Graph Diffusion Transformer (SGDiT), a flow-matching-based framework that reformulates MIMO detection as a noise-level-conditioned denoising process. Built upon a graph-based SGT backbone, SGDiT incorporates adaptive layer normalization (AdaLN) to enable stage-aware modulation across denoising steps, and adopts a prediction–loss alignment strategy tailored to discrete signals via signal-space prediction with a binary cross-entropy objective. Experimental results demonstrate that SGDiT achieves competitive performance under matched Rayleigh fading conditions and maintains strong generalization under channel distribution shifts. These results highlight the effectiveness of explicitly modeling the denoising trajectory and aligning the prediction space with the discrete nature of communication signals. Overall, the proposed framework provides an effective and practical design paradigm for neural MIMO detection, and offers a promising direction for integrating flow-based modeling with structured architectures in future communication systems.

\bibliographystyle{IEEEtran}
\bibliography{references}
\end{document}